\documentclass[12pt]{article}

\usepackage{graphicx}

\begin{document}

\title{Features of Microwave Radiation and Magnetographic
Characteristics of Solar Active Region NOAA 12242 Before the X1.8
Flare on December 20, 2014}

\author{V. E. Abramov-Maximov$^1$ \and V. N. Borovik$^1$ \and L. V. Opeikina$^2$
\and A. G. Tlatov$^1$ \and L. V. Yasnov$^3$}

\date{$^1$Central Astronomical Observatory at Pulkovo,\\ Russian Academy of Sciences, St. Petersburg, Russia\\
$^2$Special Astrophysical Observatory,\\ Russian Academy of
Sciences, Nizhnii Arkhyz, Russia\\$^3$St. Petersburg State
University,\\St. Petersburg, Russia\\-----------------\\
E-mail:beam@gaoran.ru\\-----------------\\Published
in:\\Geomagnetism and Aeronomy, 2017, Vol. 57, No. 8, p.978\\{\bf
DOI}: 10.1134/S0016793217080023}

\maketitle

\begin{abstract}
This paper continues the cycle of authors' works on the detection
of precursors of large flares (M5 and higher classes) in active
regions (ARs) of the Sun by their microwave radiation and
magnetographic characteristics. Generalization of the detected
precursors of strong flares can be used to develop methods for
their prediction. This paper presents an analysis of the
development of NOAA AR 12242, in which an X1.8 flare occurred on
December 20, 2014. The analysis is based on regular multiazimuth
and multiwavelength observations with the RATAN-600 radio
telescope in the range 1.65-10 cm with intensity and circular
polarization analysis and data from the Solar Dynamics Observatory
(SDO). It was found that a new component appeared in the AR
microwave radiation two days before the X-flare. It became
dominant in the AR the day before the flare and significantly
decreased after the flare. The use of multiazimuth observations
from RATAN-600 and observations at 1.76 cm from the Nobeyama
Radioheliograph made it possible to identify the radio source that
appeared before the X-flare with the site of the closest
convergence of opposite polarity fields near the neutral line in
the AR. It was established that the X-flare occurred 20 h after
the total gradient of the magnetic field of the entire region
calculated from SDO/HMI data reached its maximum value. Analysis
of the evolution of the microwave source that appeared before the
X-flare in AR 12242 and comparison of its parameters with the
parameters of other components of the AR microwave radiation
showed that the new source can be classified as neutral line
associated source (NLS), which were repeatedly detected by the
RATAN-600 and other radio telescopes 1-3 days before the large
flares.
\end{abstract}

\section{INTRODUCTION}

The method for predicting large flares (the most energetic
manifestations of solar activity that affect the near-Earth space
and terrestrial atmosphere (Benz, 2017)) can be based on facts
found in different solar radiation ranges that indicate the
accumulation of energy and the preparation of the active region
(AR) for a large flare (mainly, by the magnetic field
configuration and its variation), as well as by such features as
the spatiotemporal distribution of weak flares ("precursor
flares") that accompany a large flare (Gyenge et al., 2016).

In recent decades, studies have shown that flares mostly occur in
ARs with a complex magnetic structure, predominantly with a
$\delta$-configuration with a wavy neutral line and high field
gradients. The researchers have come to such conclusions by
analyzing the data of the GOES, RHESSI, SOHO/MDI, SDO (AIA and
HMI), and other space observatories. Several reviews of active
solar events with large flares of M5 and higher classes (based on
X-ray classification) based on the databases of space
observatories have recently been published. In these papers, the
analysis of individual active events is generalized, and a large
number of events with large flares are statistically analyzed in
order to identify the characteristic features in AR development,
i.e., precursors of large flares and CMEs (Schwijer et al., 2016;
Harra et al., 2016; Turiumi et al., 2017). It has been established
that, in the vast majority of cases, large flares can be predicted
by the emergence of a new magnetic flux, the presence of a
strongfield, high-gradient polarity inversion line (SHIL), motions
of magnetic fluxes, shear generation, sunspot rotation, and the
occurrence of electric currents in the AR.

Important information about the accumulation of energy and an
upcoming large flare in the AR can be provided by microwave
observations of the Sun. The first evidence of the relationship
between the features of the AR microwave radiation structure and
its flare activity were obtained in the 1970s. Thus, during
observations of the solar eclipse on March 7, 1970, a group of
Pulkovo radio astronomers found an intense, compact, and weakly
polarized radio source at 4.5 cm in the flare-active group over
the region between sunspots (Quinones et al., 1975). In 1974,
observations with the Westerbork Synthesis Radio Telescope (WSRT)
at a wavelength of 6 cm revealed a quasi-stationary bright source
that was projected on the region near the neutral line of the
photospheric magnetic field (neutral line associated source, NLS)
(Kundu et al., 1977). Subsequent observations of the three ARs
with the WSRT in May 1980 confirmed that there were intense
sources (NLSs) associated with X-ray loops over the neutral line
and that one of them had intensified several hours before a large
flare (Kundu and Alissandrakis, 1984).

From then on, radio observations of the Sun by large instruments
(the RATAN-600, Nobeyama Radioheliograph (NoRH), and Siberian
solar radio telescope (SSRT)), as well as radio observations of
solar eclipses, showed that the characteristic feature of ARs with
high flare activity is the presence of microwave NLSs. Regular
observations of the Sun with the RATAN-600, which started in 1975
in a wide microwave range, made it possible to find that microwave
sources appeared and developed quickly before large flares in ARs.
These sources were identified by the sites of the maximum gradient
of the photospheric magnetic field near the neutral line, which
often became the dominant component in the AR microwave radiation.
The characteristics of these radio sources (the steep slope of the
flux density spectrum of in the range 2-4 cm, low polarization
degree (not more than 30\%), small angular sizes, and high
brightness temperature) distinguished these sources from other
well-studied AR microwave radiation components associated with
sunspots, flocculi, and coronal condensations. That is why these
sources are called {\it peculiar sources} (see, for example,
(Akhmedov et al., 1986, 1989; Borovik et al., 1989a, b)).
Korzhavin et al. (1989) suggested that peculiar sources are
connected with strong currents (current sheets) in the atmosphere
of flare ARs.

Joint observations of a number of ARs with the WSRT and RATAN-600
at wavelengths of 2-6 cm in May 1980 showed that NLSs observed
with the WSRT and peculiar sources observed with the RATAN-600 are
the same objects (Alissandrakis et al., 1993). Subsequently,
compact intense microwave sources identified with the NL of the
photospheric magnetic field or regions near the NL were called
NLSs. Such sources are recognized as typical (often, dominant)
components of the microwave radiation (of mainly flare) ARs with a
complex magnetic structure (Shibasaki et al., 2011). The nature of
NLSs has been investigated (Yasnov, 2014).

Observations of the Sun with the NoRH at a wavelength of 1.76 cm
and with the SSRT at 5.2 cm revealed bright quasistationary
sources above the neutral line in ARs in which there were large
flares and large CMEs, in particular, those that caused the GLE
(Uralov et al., 2000; 2006; 2008). The studied NLSs at a
wavelength of 1.76 cm were identified by their location on or near
the neutral line of the vertical component of the coronal magnetic
field at the point where the horizontal component has an absolute
or local maximum. It was found that NLSs are connected with the AR
energy release sites, and it was concluded that the microwave
radiation of large flares starts under the coronal current sheet
in magnetic loops, the bases of which are in the area of motions
of strong photospheric fields.

Studies of microwave characteristics of a number of ARs with high
flare activity continued with the RATAN-600 in 2011-2015, and the
analysis of the AR characteristics by SDO data (Abramov-Maximov et
al., 2015a, b) showed that, in the studied ARs, a new (in most
cases, dominant) microwave source (NLS) was detected 1-2 days
before X class flares (according to the GOES classification) in
the AR microwave radiation, which was associated with the site of
the maximum gradient of the photospheric magnetic field near the
neutral line. The evolution of this source was similar to the
magnetic field gradient evolution. Analysis of the magnetic
measurements of the same ARs by the SDO data showed that X-class
flares occurred at sufficiently high magnetic flux levels in
sunspot groups (F $\sim$ 1022 Mx) and at a sharp increase in the
magnetic field gradient, which reflected the geometric approach of
opposite polarity sunspots.

This paper continues the cycle of authors' works on the detection
of the precursors of large flares (M5 and higher classes) in ARs
of the Sun by their microwave radiation and magnetographic
characteristics. Generalization of the detected precursors of
strong flares can be used to develop methods for their prediction.
Analysis of the development of AR NOAA 12242, in which an X1.8
flare occurred on December 20, 2014, is presented. The analysis is
based on regular observations of the Sun in the microwave range
with the RATAN-600 radio telescope. Magnetographic characteristics
of ARs were studied based on SDO/HMI data.

\section{RATAN-600 OBSERVATIONS OF THE SUN}

In this paper, we used the results of daily multiazimuth
spectral-polarization multiwavelength observations of the Sun
carried out with the RATAN-600 radio telescope in the selected
wavelength range of (1.8-4.0) cm. Observations were conducted with
a multioctave high-resolution spectral-polarization complex (Bogod
et al., 2011).

The knife-edge beam at a wavelength of 2.0 cm (HPBW) is
$17''\times13'$. In December 2014, observations of the Sun on
RATAN-600 were carried out in multiazimuth mode, i.e., 31
observations were conducted daily in different azimuths in the
time interval of 06:50-11:30 UT, while the Sun scanning angle by
the antenna beam varied within $\pm$14.5 deg. (Bogod et al.,
2004). This observation method made it possible both to
investigate the dynamics of the development of individual
microwave radiation components of AR 12242 for more than 4 h and
to carry out their two-dimensional identification with parts of
the AR.

An example of the results of observations of the Sun with the
RATAN-600 on December 18, 2014, in one of the 31 azimuths is given
in Fig. 1. The considered AR 12242 is in the right part of the
disk, and AR 12241 is in the left part. In Fig. 1, solid lines
indicate 7 one-dimensional scans in the intensity channel (Stokes
parameter I) in the order of increasing wavelengths given on the
left (in cm). The dashed lines show scans in the polarization
channel (Stokes parameter V) in two wavelengths (2.0 and 2.2 cm).
The scans are superimposed on the image of the Sun (SDO). AR 12241
was used to control the referencing of scans to the image of the
Sun.

The parameters of individual components of the radio emission of
the considered AR 12242 were estimated by the Gauss analysis by
the method of processing of solar scans obtained with the RATAN-
600 (Akhmedov et al., 1987) using the WorkScan software (Garaimov,
1997). The radio source fluxes identified on one-dimensional solar
scans were calibrated with the use of the results of observations
of standard objects (the Moon and the Crab Nebula) with the
RATAN-600 and also with an accounting of the data of the world
solar patrol.

\section{RESULTS OF OBSERVATION OF AR 12242
WITH THE RATAN-600}

AR 12242 was first detected on the disk on December 14, 2014, as a
group of sunspots with an area of 100 msh (millionths of the solar
hemisphere) (according The Preliminary Report and Forecast of
Solar Geophysical Data,
ftp://ftp.ngdc.noaa.gov/STP/swpc\_products/weekly\_reports/PRFs\_of\_SGD).
As the group moved by the disk, it developed; its area increased
and reached a maximum value of 1080 msh on December 19. From
December 15 to 17, a number of C-class flares and two M-class
flares were detected in it (one of them was a large M8.7 flare on
December 17 (peak at 04:42 UT)). Unfortunately, it is impossible
to identify the precursors of this flare because of the AR
closeness to the eastern limb in the period until December 17.
Therefore, in this paper, the task is to investigate the evolution
of the AR 12242 microwave radiation and its magnetographic
characteristics before the large X1.8 flare (beginning at 00:11
UT, peak at 00:28 UT, and end at 00:55 UT) for December 17-20.

Figure 2 shows fragments of solar scans (RATAN-600) at several
wavelengths identified with the AR 12242 image in white light and
magnetograms (SDO/HMI) for December 16220, 2014. It can be seen
that, two days before the X-flare, new components appeared in the
AR microwave radiation structure, one of which (designated as 2)
became the dominant component a day before the flare. After the
flare, it noticeably decreased.

The identification of AR microwave radiation components based on
observations with the RATAN-600 in different azimuths with
different position angles of the Sun showed that source 2 is
uniquely identified with the region near the neutral line, where
the negative polarity field came close to the previously existing
positive polarity field (in Fig. 2 and Fig. 4b, it is marked by
arrow 2). Source 3 is identified by multiazimuth observations with
a region covering the southern part of the large negative polarity
sunspot in the tail part of the group and the region near the
neutral line, where the negative polarity field also came close to
the positive polarity field (indicated in Fig. 2 and Fig. 4b by
arrow 3).

Solar radio images at a wavelength of 1.76 cm obtained with the
Nobeyama Radioheliograph were used for more reliable
identification of microwave sources with a magnetogram. In Fig.
3a, radio images of NOAA ARs 12241 and 12242 at 1.76 cm on
December 19 are superimposed on the magnetogram (SDO/HMI). The
white solid isolines show the intensity (Stokes parameter I), and
the dashed lines indicate circular polarization (Stokes parameter
V): the white lines refer to positive polarization, and black
refer to negative polarization. The leading (head) sunspot of AR
12241 was used to control the coordinate referencing of the radio
image and the magnetogram.

Figure 3a shows that the radiation maximum at 1.76 cm coincides
with region 2 in Fig. 2 near the neutral line, i.e., with the site
of maximum convergence of the magnetic fields with the opposite
polarity sign. Region 3 near the neutral line to the south of the
tail sunspot (see Fig. 2), where opposite polarity fields also
converged, is also located in the high radiation intensity region
at a wavelength of 1.76 cm.

Thus, the identification of intense microwave sources detected on
the RATAN-600 (one day before the flare) with the sites of
convergence of opposite polarity fields near the neutral line
agrees with the results of observations on the NoRH at 1.76 cm.

It is important to note that the X1.8 flare (beginning at 00:11 UT
and peak at 00:28 UT) began to develop on December 20 exactly in
this region of the neutral line of the photospheric magnetic
field, where the fields of opposite signs converge. This can be
seen in Fig. 3b, where the image of the AR 12242 in the line 1600
Å (SDO) is given at 00:17:29 UT on December 20.

\section{MAGNETOGRAPHIC INVESTIGATIONS
OF AR 12242}

This paper presents the results of the analysis of the structure
and magnetic characteristics of sunspots in AR 12242 for December
17–21, 2014, according to SDO data. The SDO/HMI data was analyzed
with an accumulation time of 45 s at 00:00, 05:00, 10:00, 15:00,
and 20:00 UT. Automatic identifying of sunspots and measurement of
the magnetic flux in them by SDO/HMI observations in the continuum
and in magnetographic observations was used (Tlatov et al., 2014).

The method for calculating the index of solar flares used in this
paper can be described in the following steps. At the first step,
the boundaries of ARs with different polarities were determined.
For this purpose, structures with a magnetic field above the
threshold values ($\pm$500 G) were selected on the HMI/SDO
magnetograms. The effective distance between the boundaries of the
regions was then determined.

In order to simplify computational algorithms, only distance data
in the horizontal and vertical directions were taken into account.
At the first step of the algorithm, vertical scanning was carried
out. Thus, when scanning the current column, the positive region
boundary (point $i$) was found. Next, the total positive polarity
magnetic flux $\Phi^l_{pos}$ was determined. The pixels were
scanned from point $i$ by the column until the first boundary
point of the negative region (point $j$) was detected. After point
$j$ was found, it was taken into account that this point belonged
to the negative region, and the flux of this region was
$\Phi^k_{neg}$. If no such point was detected, then the given
point $i$ was no longer considered.

Next, the distance between the points $i$ and $j$
$d_{ij}=\sqrt{(\Theta_i-\Theta_j)^2+(\varphi_i-\varphi_j)^2}$ was
determined, where $\Theta$ and $\varphi$ are the latitude and
longitude of the points $i$ and $j$. If the distance $d_{ij}$ was
greater than the threshold value $d_{max}=5deg$, then this pair
was not considered further. The current gradient
$grad_{ij}=(\Phi^l_{pos}+\Phi^k_{neg})/d_{ij}$ was then
calculated. The index $grad_{ij}$ was summed to determine the
total index of bipole moments (flare index).

In Fig. 2, magnetograms show how the configuration of the magnetic
field in AR 12242 changed from December 16 to 20: on December 18,
a negative polarity field appeared to the south of the tail
sunspot near the positive polarity field that existed on the
previous day. Simultaneously, a negative polarity field westwardly
approached the positive polarity field along the neutral line. The
next day, on December 19, the opposite polarity fields got as
close as possible at two sites near the neutral line (at points 2
and 3 marked by the corresponding arrows). On December 20, after
the X1.8 flare, the negative polarity field at point 2 departed
from the positive polarity field, while at point 3 the negative
polarity field remained near the positive polarity field (see also
Fig. 4c).

The magnetographic characteristics of the entire AR 12242 and
specific regions 2 and 3 near the neutral line calculated based on
the SDO/HMI data as described above are shown in Fig. 5, where the
graphs of development of the total bipole moment index INDEX for
the whole AR and magnitudes of gradients GRAD in separate regions
2 and 3 are given. A similar behavior in the development of both
the total index and the gradients in both regions for the period
from December 17 to 20 can be observed: the total index and
gradients increase two days before the X-flare, and, after they
pass a maximum (15-20 h before the flare), there is a decline
before the X-flare.

In Fig. 4, the crosses on the AR magnetograms indicate the
positions of the maximum gradient at times 02:29:39 and 02:59:09
UT. Calculations carried out with an interval of 10–30 min showed
that, for the entire considered period starting from December 18,
the maximum gradient in different time intervals for a long time
was either in region 2 or in region 3. However, on December 19,
the maximum gradient sites were alternately in regions 2 or 3,
which is evident in the above examples on the magnetograms in Fig.
4a and 4b obtained with an interval of 30 min. After the X-flare
on December 20, the gradient maximum was stably in region 3, and
the gradient at point 2 decreased significantly. The negative
polarity field moved away from the positive polarity field, and
its intensity decreased by 20\%.

\section{ANALYSIS OF MICROWAVE OBSERVATIONS OF AR 12242}

Let us consider the evolution of individual components of the
microwave radiation of AR 12242 for December 17-20. Figure 5 shows
changes in the intensity of individual components of AR microwave
radiation in four days. Figures 5a and 5b show the intensities at
2.36 cm and 3.45 cm of microwave sources 2 (panel (a)) and 3
(panel (b)) expressed in antenna temperatures Ta over the
observation periods with the RATAN-600 (daily in time intervals
from 07:00 to 11:30 UT). In Fig. 5c, the evolution of source 3 at
the wavelengths of 2.2, 2.36, 2.75, and 3.45 cm for December 18 is
given in more detail. Here, in the upper part of the figure, the
bolder line shows the change in the integral flux of the radio
emission from the Sun at a wavelength of 3.2 cm according to
observations at Mountain (Kislovodsk) Astronomical Station of the
Pulkovo observatory. Vertical lines indicate the moments of radio
observations in different azimuths. The thicker line corresponds
to the observation time at the local noon. Figure 5c clearly shows
a high correlation of variations in the intensity of the sources
at all the given waves with variations in the integral flux of the
Sun, which is due to microwave bursts associated with flares in AR
12242. A similar conclusion was drawn via comparison of variations
in the intensity of sources 2 and 3 during observations with the
RATAN-600 with variations in the integral flux of the Sun on other
days.

Comparison of the evolution of the intensity of the sources that
developed before the X-flare with the development of the magnetic
field gradient in the regions near the neutral line with which
they are identified (at points 2 and 3 in Fig. 2) is of particular
interest. As can be seen from Fig. 5a, the evolution of source 2
(in particular, at 3.45 cm wavelength) reflects the evolution of
the gradient 2 for December 18-20. Simultaneously, the intensity
of source 3 changed insignificantly for the same period. This can
be explained by the fact that the radiation of the sunspot
microwave source associated with the tail sunspot of the group
and, possibly, of the source generated near the neutral line
associated with gradient 3 simultaneously entered a knife-edge
antenna pattern. Apparently, the sunspot source radiation
dominated, and it was quite stable (if we take into account that
the field intensity in the tail spot for December 18-20 did not
change by more than 10\%).

The characteristics of the individual components of the microwave
radiation of AR 12242 obtained from observations with the
RATAN-600 were analyzed. Particular attention was paid to a
comparison of the parameters of sources 2 and 3, the most intense
microwave sources in the AR emission that were detected before the
X1.8 flare (see Fig. 2).

Figure 6a shows the spectra of source flux densities for different
observation days. By comparing the spectra of the sources obtained
on December 19 (for the times when there were no microwave bursts
on the Sun), two main differences can be noted:

(a) In the short-wave part (at waves shorter than 2.5 cm), the
source 2 spectrum slope is greater than that of source 3.

(b) In the case of the complete coincidence of the fluxes of both
sources in the range of 2.5-3.2 cm, a difference in the fluxes (by
60\%) on longer waves (about 4 cm) can be seen. If the angular
sizes of both sources are close (as can be seen from the structure
of the AR on short wavelengths), it can be assumed that the
brightness temperatures of source 2 at wavelengths of about 4 cm
are higher than those of source 3. However, it is impossible to
reliably determine the brightness temperature of sources at
wavelengths of about 4 cm because of the insufficient antenna
resolution and, accordingly, because it is impossible to determine
unambiguously the sizes of the sources. In this case, it can be
stated that the brightness temperatures of the sources at
wavelengths shorter than 3 cm are close and that they are about
$(3-4)\times10^6$ K at wavelengths longer than 3 cm.

Table 1 shows the spectral indices of the sources determined in
the wavelength range of 1.8-2.5 cm according to observations on
different days.

The spectrum of source 2 has the largest spectral index. The
spectrum of source 3 has a smaller spectral index, which can be
explained by the fact that the flux of source 3 was composed of
the dominant flux of the sunspot source associated with a large
tail sunspot and probably an existing source associated with the
formed magnetic field gradient near the neutral line. However,
when observing with a knife-edge antenna pattern, one cannot
separate the radiation from these sources.

The spectrum of source 3 for December 17 reflects the total
radiation of only two sunspot sources associated with large
sunspots in the tail of the group (as seen in Fig. 2, the magnetic
field gradient near the neutral line has not yet formed on this
day). On December 20, the spectrum of the weak source remaining
after the X-flare also has a low spectral index. Note that the
gradient decreased significantly after the X-flare in this region
near the neutral line as compared to what it was the day before
the flare on December 19.

A comparison of the spectra of polarization degrees for sources 2
and 3 is presented in Fig. 5b. It can be seen that the
polarization degree of source 2 is smaller than polarization
degrees of source 3.

Thus, the main features by which the characteristics of microwave
source 2 turned out to be different from these of source 3, in
which the sunspot radiation prevails, were as follows: a steeper
spectral slope (large spectral index) in the short-wavelength part
of the centimeter range, small angular sizes comparable with the
size of sunspot source, and a smaller polarization degree. Taking
into account the fact that source 2 is identified with the site of
the closest convergence of opposite polarity fields, i.e., with
the site of the maximum gradient of the photospheric magnetic
field near the neutral line, all of the above radio
characteristics make it possible to attribute it to NLSs, which
were repeatedly observed before large flares.

\section{CONCLUSIONS}

The results were presented for a study of the evolution of the
microwave radiation and the magnetographic characteristics of NOAA
AR 12242, in which there was a X1.8 flare on December 20, 2014
(according to the GOES classification). The analysis was carried
out to identify the precursors of a large flare in the microwave
range based on daily multiwavelength spectral-polarization
observations of the Sun in the 1.65–10 cm range with the RATAN-600
radio telescope. SDO/HMI data were used to analyze the AR
magnetographic characteristics.

Two days before the flare, the development of a new microwave
source was detected; it became the dominant component in the AR
radio emission the day before the X-flare and significantly
decreased after the flare. It was found that the new source was
identified with the site of the closest convergence of opposite
polarity fields near the neutral line of the photospheric magnetic
field. Magnetographic studies of AR 12242 showed that, during the
development of the group, opposite polarity fields converged in
the AR and strong gradients were formed near the neutral line,
i.e., two days before the flare, a so-called {it SHIL}
(strong-field, high-gradient polarity inversion line) formed in
the AR that is now recognized by many researchers as the clearest
AR precursor for a large flare (Schwijer et al., 2016; Harra et
al., 2016; Turiumi et al., 2017).

The study of the evolution of the total gradient (flare index) of
AR 12242 showed that, before the X-flare, the index first
increased and then, after the maximum was passed (20 h before the
flare), decreased before the X flare. Note that a similar
conclusion was made by Korsos et al. (2014, 2015), who analyzed
the parameters of sunspot groups in a number of flare regions and
found that there is usually a sharp increase in the magnetic field
gradient before large flares that reaches a high maximum, followed
by its decrease just before the flare.

The identification of a microwave source that developed before the
large flare with the site of the maximum gradient of the AR
photospheric field near the neutral line and the parameters of its
radio emission makes it possible to classify it as an NLS that was
detected before flares in early solar observations with the
RATAN-600, and were revealed as precursors of large flares in the
analysis of a number of active events in 2011-2015 based on data
from the RATAN-600 and SDO (Abramov-Maximov et al., 2015a,b).

The results make it possible to conclude that regular radio
observations of the Sun in the microwave range can be used to
identify precursors of large flares, based on which methods for
their prediction can be developed.

\section{ACKNOWLEDGMENTS}

This study was partially supported by the Program of the Presidium
of the Russian Academy of Sciences.

We thank the staff of the Special Astrophysical Observatory of the
Russian Academy of Sciences for observations of the Sun with the
RATAN-600 and A.D. Shramko for observations of the Sun at Mountain
Astronomical Station of the Pulkovo observatory. We are grateful
to the SDO team for the HMI(SDO) observational data. This work was
performed using of the Nobeyama Radioheliograph operated by the
International Consortium for the Continued Operation of Nobeyama
Radioheliograph (ICCON). ICCON consists of the ISEE/Nagoya
University, NAOC, KASI, NICT, and GSFC/NASA. L.V. Yasnov
acknowledges the support of the Russian Foundation for Basic
Research, project no. 16-02-00254. V.E. Abramov-Maximov
acknowledges the support of the Russian Science Foundation,
project no. 16-12-10448. We are grateful to the reviewer for
helpful remarks.

\begin{figure}
\centerline{\includegraphics[width=10cm]{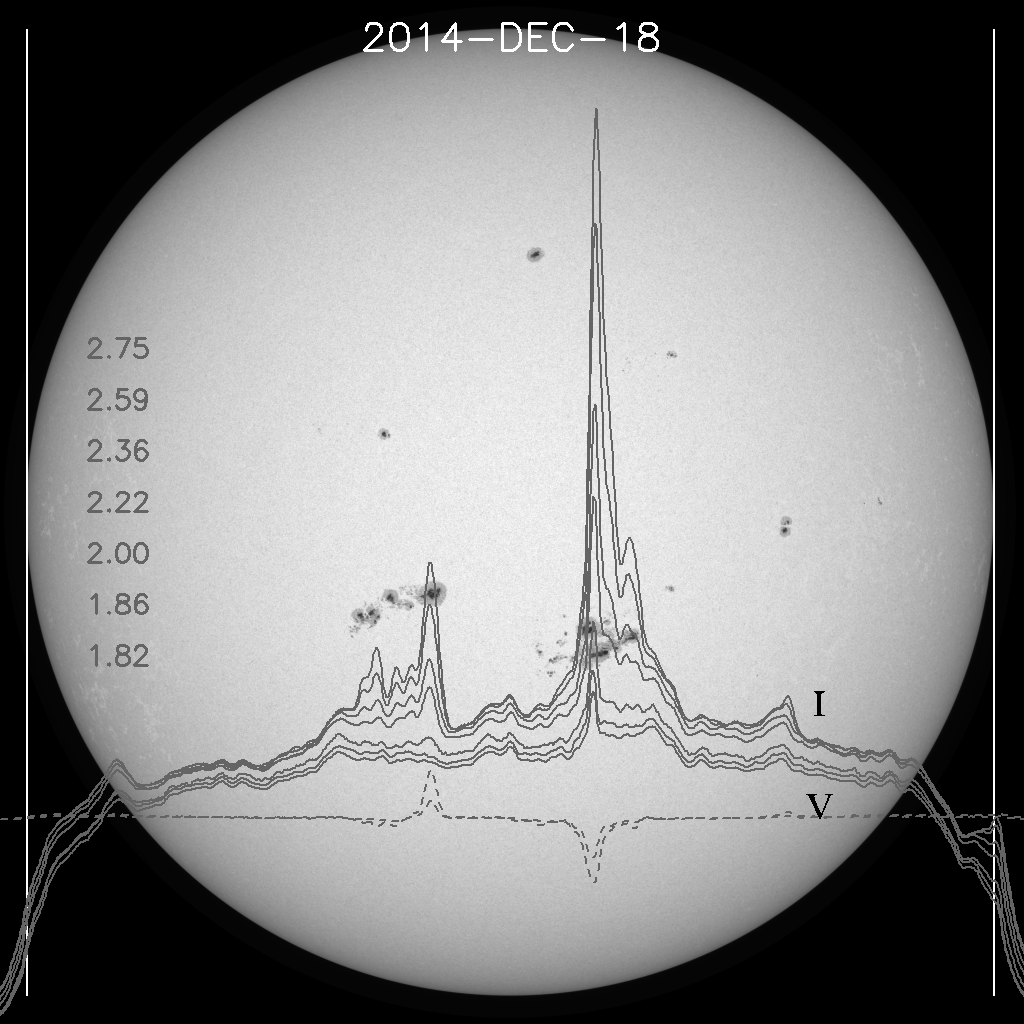}}
\caption{Example of the results of one observation of the Sun with
the RATAN-600 for Dec. 18, 2014. Solid lines indicate
one-dimensional scans in the intensity channel (Stokes parameter
I); the wavelengths in cm are given on the left. Scans are
arranged in the order of increasing wavelengths. Dashed lines show
scans in the polarization channel (Stokes parameter V) in two
wavelengths (2.0 and 2.2 cm). The scans are superimposed on the
image of the Sun (SDO).}
\end{figure}

\begin{figure}
\centerline{\includegraphics[width=15cm]{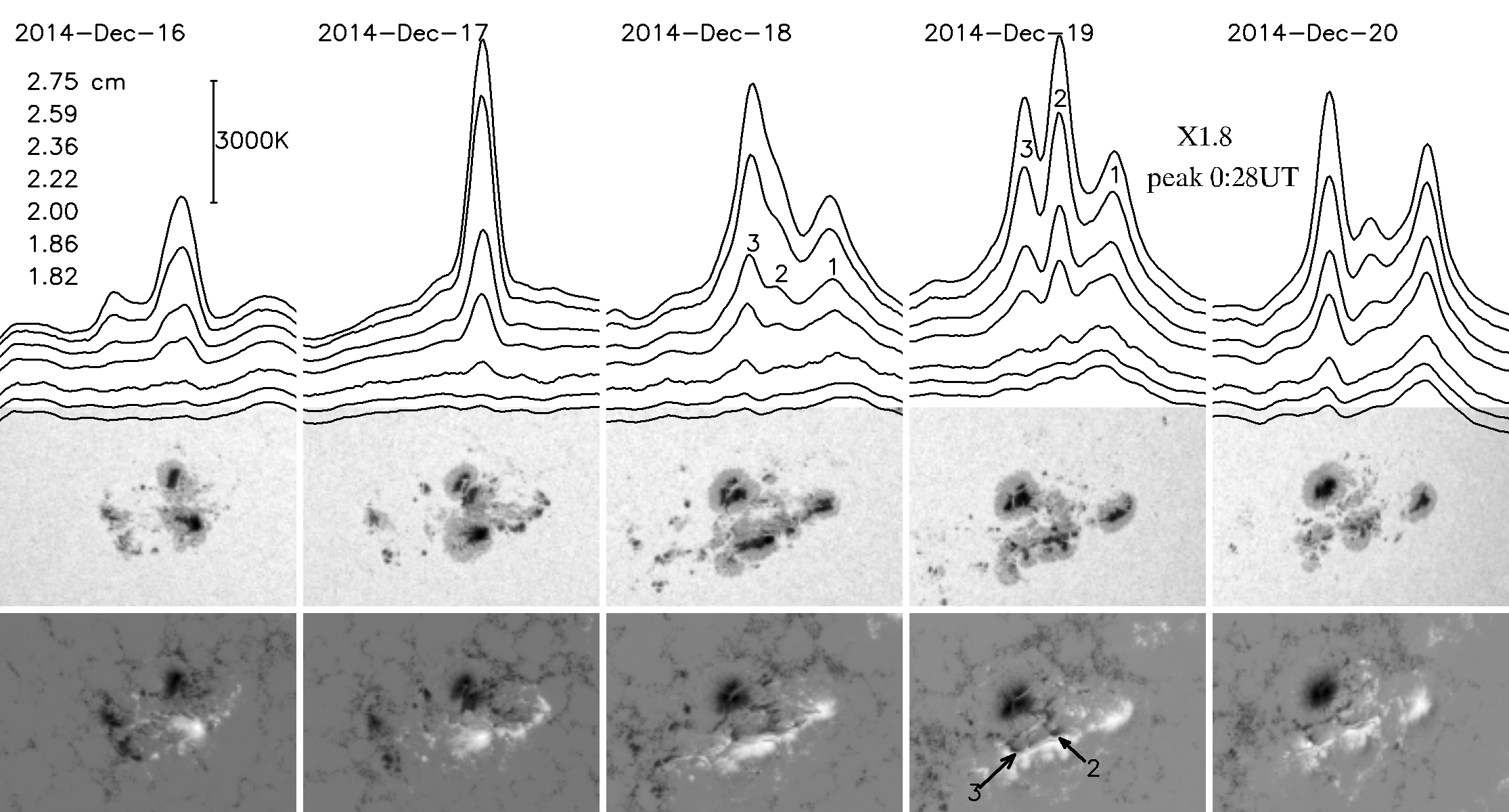}}
\caption{Fragments of one-dimensional solar scans in the microwave
range in the intensity channel (Stokes parameter I) obtained with
the RATAN-600. The scans are arranged in the order of increasing
wavelengths from the bottom up. The wavelengths in cm are given in
the left panel. The vertical fragment shows the scale in antenna
temperatures (K). AR NOAA 12242 images in the continuum and
magnetograms according to SDO/HMI data are given under the scans.
The AR microwave radiation components are marked on the panels for
Dec. 18 and 19, 2014. In the magnetogram for Dec. 19, the arrows
indicate two regions of the maximum magnetic field gradient at the
level of the photosphere.}
\end{figure}

\begin{figure}
\centerline{\includegraphics[width=10cm]{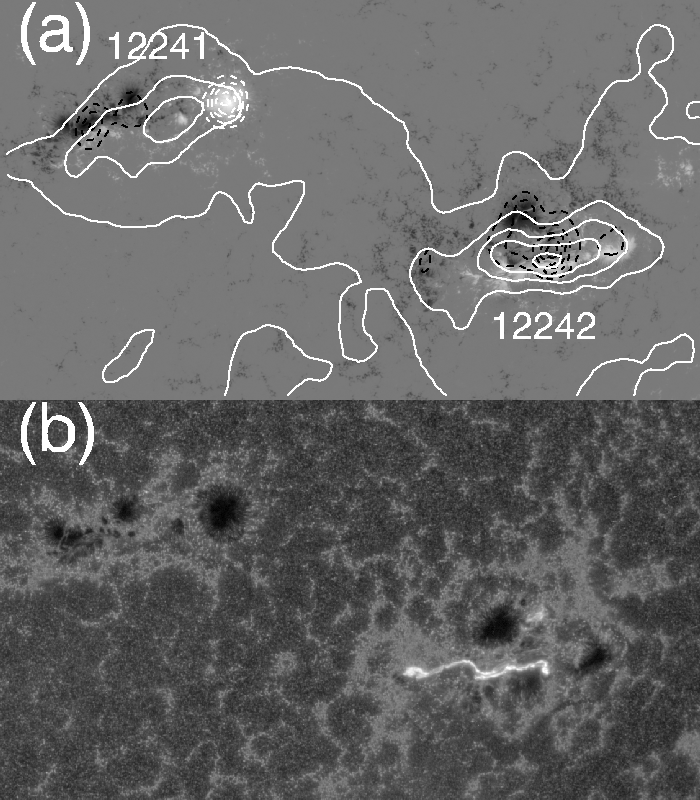}}
\caption{(a) Image of ARs NOAA 12241 and 12242 at a wavelength of
1.76 cm obtained with the Nobeyama Radioheliograph on Dec. 19,
2014, at 02:44 UT, superimposed on the magnetogram (SDO/HMI).
White solid isolines show the intensity (Stokes parameter I), and
dashed lines indicate circular polarization (Stokes parameter V):
white lines refer to positive polarization, and black lines refer
to negative polarization. The leading sunspot of AR 12241 was used
to control the coordinate referencing of the radio image and the
magnetogram. (b) Image of ARs NOAA 12241 and 12242 in the line
1600~\AA{} (SDO/AIA) on Dec. 20, 2014, at 00:17 UT.}
\end{figure}

\begin{figure}
\centerline{\includegraphics[width=15cm]{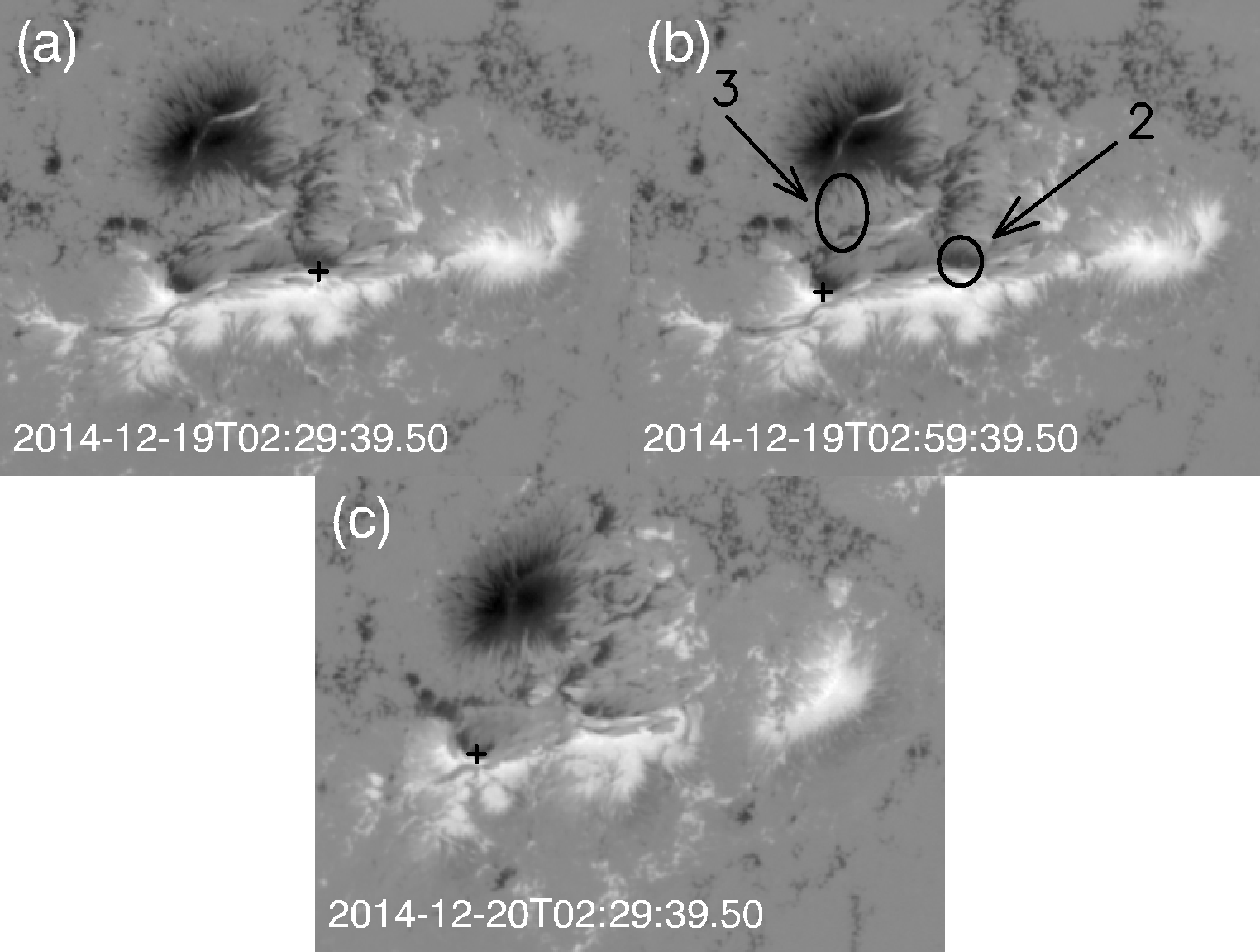}}
\caption{Magnetograms of the AR NOAA 12242 (SDO/HMI). The cross
shows the positions of the maximum gradient of the magnetic field
at the level of the photosphere. The panel (b) shows the areas
above which the microwave sources 2 and 3 are located. Positions
of the regions are determined from observations in different
azimuths with different positional angles.}
\end{figure}

\begin{figure}
\centerline{\includegraphics[width=15cm]{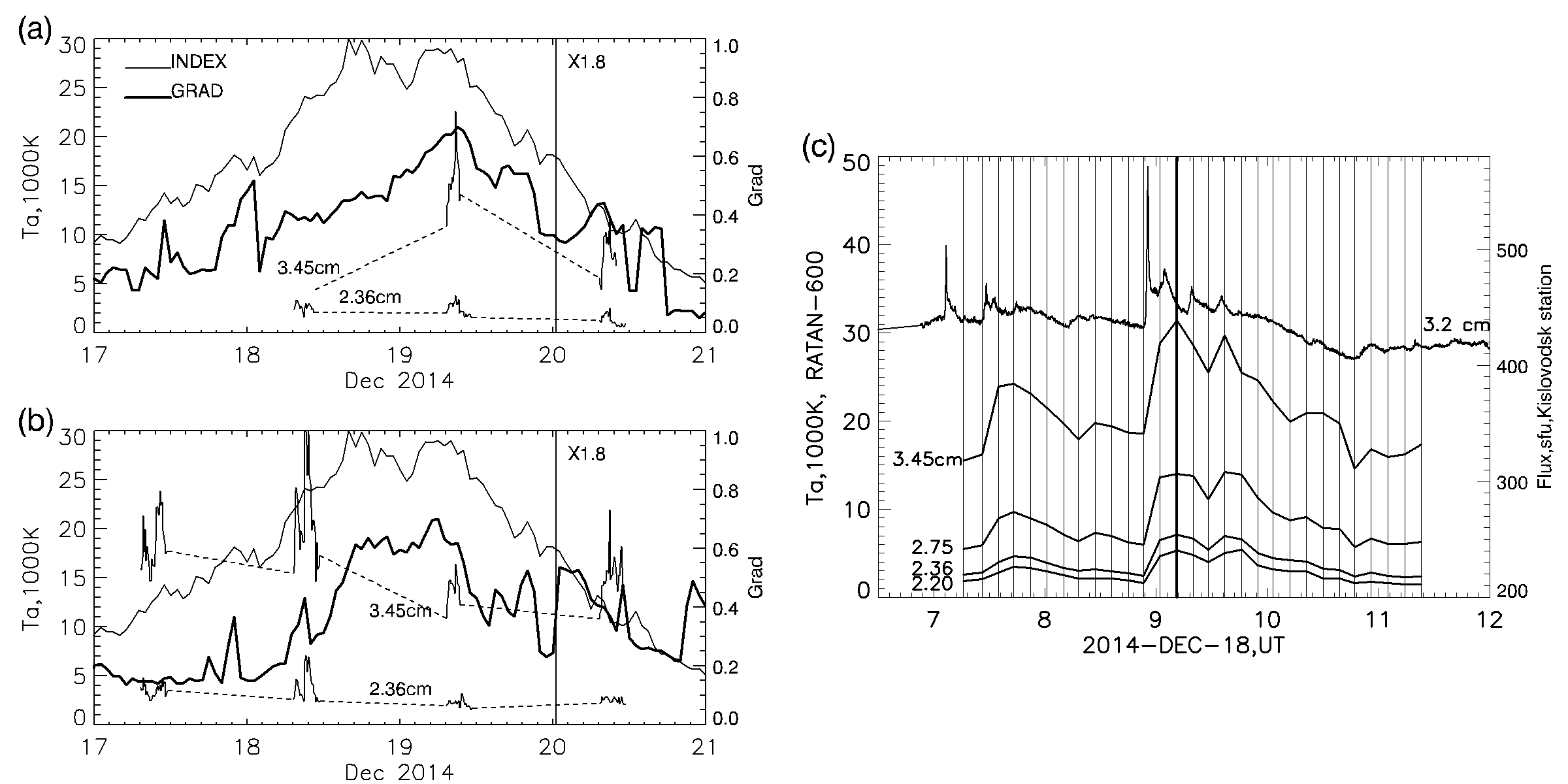}}
\caption{Evolution of microwave radiation and the solar flare
index of the AR NOAA 12242. (a) The total index of the bipole
moments (INDEX) and the gradient (GRAD) at point 2 (see Fig. 2)
(right ordinate axis, arbitrary units), and the intensity of
microwave radiation in antenna temperatures (left ordinate axis)
of source 2 (see Fig. 2) at 2.36 and 3.45 cm. The vertical line
shows the X1.8 flare moment. (b) Same as in panel (a) for source
3. (c) Microwave radiation of source 3 of AR NOAA 12242 on Dec.
18, 2014, based on multiazimuth observations with the RATAN-600 at
2.2, 2.36, 2.75, and 3.45 cm (left ordinate axis) and the total
flux of the radio emission of the Sun at a wavelength of 3.2 cm
according to observations at Mountain (Kislovodsk) Astronomical
Station of the Pulkovo observatory (right ordinate axis). Vertical
lines indicate the moments of observations with the RATAN-600. The
thicker line marks the time of observation at the local noon.}
\end{figure}

\begin{figure}
\centerline{\includegraphics[width=15cm]{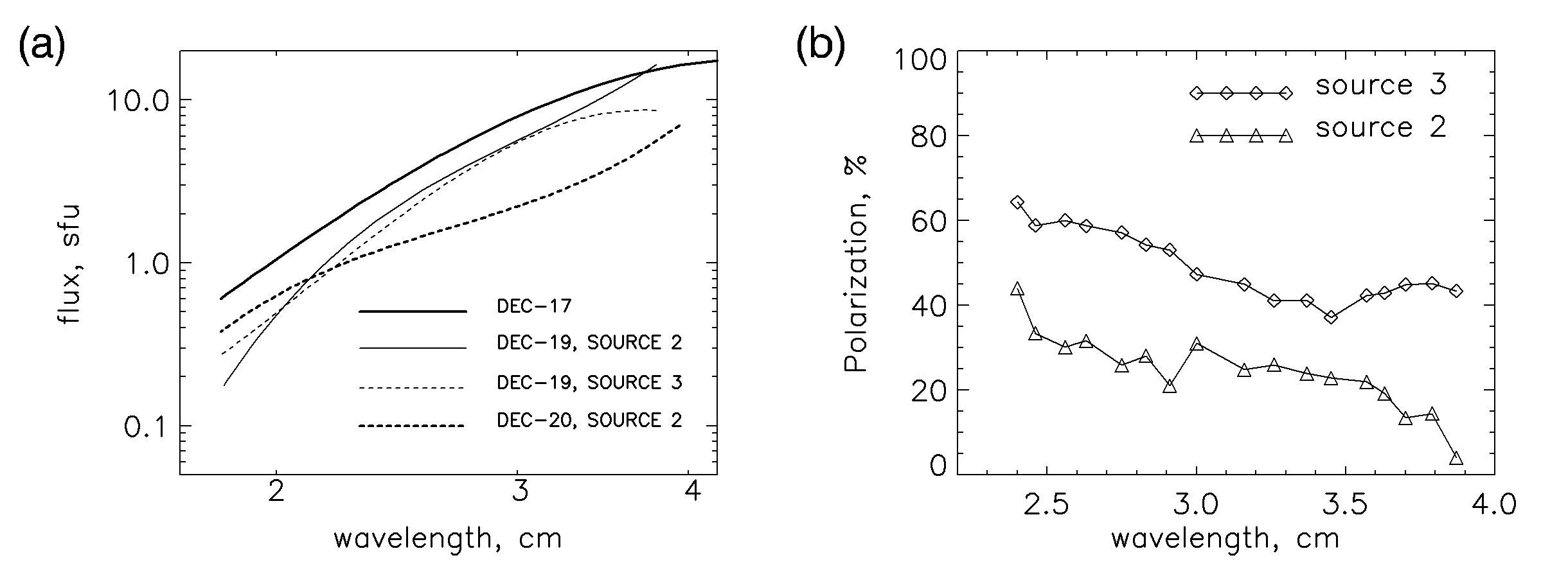}}
\caption{(a) Spectra of flux densities of different components of
the microwave radiation of AR NOAA 12242 from observations with
the RATAN-600 for Dec. 17--20, 2014. The abscissa is the
wavelengths (cm), and the ordinate axis is the flux density (sfu).
(b) The spectra of polarization degrees of sources 2 and 3 (see
Fig. 2) for Dec. 19, 2014, from observations with the RATAN-600.}
\end{figure}

\begin{table}[p]
\centering
\begin{tabular}{c|c}
\hline
  Date (December 2014) & Spectral index \\\hline
  17 & 5.6 \\
  19, source 2 & 8.5 \\
  19, source 3 & 6.5 \\
  20, source 2 & 3.5 \\\hline
\end{tabular}
  \caption{Spectral indices of microwave sources of AR
NOAA 12242}
\end{table}


\begin{thebibliography}{}

\bibitem{Abramov2015a}
Abramov-Maximov V.E., Borovik V.N., Opeikina L.V., Tlatov A.G.
Dynamics of Microwave Sources Associated with the Neutral Line and
the Magnetic-Field Parameters of Sunspots as a Factor in
Predicting Large Flares // Solar Phys. V. 290. P. 53-77. 2015à.

\bibitem{Abramov2015b}
Abramov-Maximov V.E., Borovik V.N., Opeikina L.V., Tlatov A.G.
Precursors of the solar X flare on march 29, 2014, in the active
region NOAA 12017 based on microwave radiation and magnetographic
data // Geomagnetism and Aeronomy. V. 55. Issue 8. P 1097-1103.
2015b.

\bibitem{Akhmedov1986}
Akhmedov Sh.B., Borovik V.N., Gelfreikh G.B., Bogod V.M.,
Korzhavin A.N., Petrov Z.E., Dikij V.N., Lang K., Willson R.
Structure of a solar active region from RATAN-600 and very large
array observations // Astrophys.J. V. 301. P. 460-464. 1986.

\bibitem{Akhmedov1987}
Akhmedov Sh.B., Bogod, V.M., Borovik, V.N. et al. Structure of
active regions on the Sun from VLA and RATAN-600 observations in
July 1982 // Astrofiz. Issled. (Izv. SAO). V. 25. P. 105-134.
1987.

\bibitem{Akhmedov1989}
Akhmedov Sh.B., Borovik V.N., Gelfreikh G.B., Bogod V.M.,
Korzhavin A.N., Petrov Z.E., Hofmann A., Bachmann G. Cooperative
magnetographic and radioastronomical investigations of the active
solar region AR 3804 in July 1982 // Astrofiz. Issled. (Izvest.
SAO). V. 28. P. 111-122. 1989.

\bibitem{Alissandrakis1993}
Alissandrakis C.E., Gelfreikh G.B., Borovik V.N., Korzhavin A.N.,
Bogod V.M., Nindos A., Kundu M.R. Spectral observations of active
region sources with RATAN-600 and WSRT // Astron. Astrophys. V.
270. P. 509-515. 1993.

\bibitem{Benz2017}
Benz A. O. Flare Observations // Living Rev. Sol. Phys. V. 5. P.
1-59. 2017.

\bibitem{Bogod2004}
Bogod V.M., Zhekanis G.N., Mingaliev M.G., Tokhchukova S.Kh.
Multi-azimuth regime of ob-servations at the RATAN-600 southern
sector with periscope reflector // Radiophys. Quantum Electron. V.
47. No. 4. P. 227-237. 2004.

\bibitem{Bogod2011}
Bogod V.M., Alesin A.M., Pervakov A.A. RATAN-600 radio telescope
in the 24th solar activity cycle. II. Multioctave spectral and
polarization high resolution solar research system // Astrophys.
Bull. V. 66. No. 2. P. 205-214. 2011.

\bibitem{Borovik1989a}
Borovik V.N., Drake N.A., Golovko, A.A. Evolution of the magnetic
flux in the flaring active region based on optical and radio
observations // In: Teplitskaya, R.B. Ed. Solar Magnetic Fields
and Corona, Proceedings of the XIII Consultation Meeting on Solar
Physics, 26 September - 2 October, 1988, Odessa. Nauka. Siberion
Division. Novosibirsk. V. 2. P. 162-166. 1989a.

\bibitem{Borovik1989b}
Borovik V.N., Drake N.A., Korzhavin A.N., Plotnikov V.M. The
evolution and structure of the flare-active region HR 16 631
(February 1980) based on RATAN-600 observations
// Kinemat. Fiz. Nebesnyh Tel. V. 5. P. 63-67. 1989b.

\bibitem{Garaimov1997}
Garaimov V.I. Processing of one-dimensional data vector arrays in
Windows OS, WorkScan version 2.3 // Preprint no. 127T. Nizhnii
Arkhyz: SAO. 1997.

\bibitem{Gyenge2016}
Gyenge N., Ballai I., Baranyi T. Statistical study of
spatio-temporal distribution of precursor so-lar flares associated
with major flares // MNRAS. V. 459. P. 3532-3539. 2016.

\bibitem{Harra2016}
Harra L.K. et al. The Characteristics of Solar X-Class Flares and
CMEs: A Paradigm for Stellar Superflares and Eruptions? // Solar
Phys. V. 291. P. 1761-1782. 2016.

\bibitem{Korsos2014}
Korsos M.B., Baranyi, T., Ludmany A. Pre-flare Dynamics of Sunspot
Groups
// Astrophys .J. V. 789. Id. 107. 7 P. 2014.

\bibitem{Korsos2015}
Korsos M.B., Gyenge N., Baranyi T., Ludmany A. Dynamic Precursors
of Flares in Active Re-gion NOAA 10486 // Journal of Astrophysics
and Astronomy. V. 36. Issue 1. P.111-121. 2015.

\bibitem{Korzhavin1989}
Korzhavin, A.N., Gelfreikh, G.B., Vatrushin, S.M. Peculiar sources
of solar radio emission and their possible interpretation // In:
Teplitskaya, R.B. Ed. Solar Magnetic Fields and Corona,
Pro-ceedings of the XIII Consultation Meeting on Solar Physics, 26
September - 2 October, 1988, Odessa. Nauka. Siberion Division.
Novosibirsk. V. 2 P. 119-124. 1989.

\bibitem{Kundu1977}
Kundu M.R., Alissandrakis C.E., Bregman J.D., Hin A.C. 6
centimeter observations of solar ac-tive regions with 6 SEC
resolution // Astrophys.J. V. 213. P. 278-295. 1977.

\bibitem{Kundu1984}
Kundu M.R., Alissandrakis C.E. Structure and polarization of
active region microwave emission. // Solar Phys. V. 94. P.
249-283. 1984.

\bibitem{Quinones1975}
Quinones J.A., Korzhavin A.N., Peterova N.G., Santos J.
Observations of the solar eclipse on March 7, 1970 with the
polarimeter of the Havana Radio Astronomical Station at 4.5 cm. //
Soln. Dannye. ¹3. P.87-96. 1975.

\bibitem{Schrijver2016}
Schrijver C.J. The nonpotentiality of coronae of solar active
regions, the dynamics of the surface magnetic field, and the
potential for large flares
// Astrophys.J. V. 820. Id. 103. 2016.

\bibitem{Shibasaki2011}
Shibasaki K., Alissandrakis C.E., Pohjolainen S. Radio emission of
the quite Sun and active re-gions // Solar Phys. V. 273. P.
309-337. 2011.

\bibitem{Tlatov2014}
Tlatov A.G., Vasil'eva V.V., Makarova P.A., Otkidychev P.A.
Applying an automatic image processing method to synoptic
observations // Sol. Phys. V. 289. P. 1403-1413. 2014.

\bibitem{Toriumi2017}
Toriumi S., Schrijver C.J., Harra L.K., Hudson H., Nagashima K.
Magnetic properties of solar active regions that govern large
solar flares and eruptions // Astrophys.J. V. 834. Id. 56. 2017.

\bibitem{Uralov2000}
Uralov A.M., Nakajima H., Zandanov V.G., Grechnev V.V.
Current-sheet-associated radio sources and development of the
magnetosphere of an active region revealed from 17 GHz and Yohkoh
data // Solar Phys. V. 197. P. 275-312. 2000.

\bibitem{Uralov2006}
Uralov A.M., Rudenko G.V., Rudenko I.G. 17GHz Neutral Line
Associated Sources: Birth, Mo-tion, and Projection Effect //
Publications of the Astronomical Society of Japan. V. 58. No.1. P.
21-28. 2006.

\bibitem{Uralov2008}
Uralov A.M., Grechnev V.V., Rudenko G.V., Rudenko I.G., Nakajima
H. Microwave Neutral Line Associated Source and a Current Sheet //
Solar Phys. V. 249. P. 315-335. 2008.

\bibitem{Yasnov2014}
Yasnov L.V. On the Nature of Neutral-Line-Associated Radio Sources
// Solar Phys. V. 289. P.1215-1225. 2014.

\end{thebibliography}
\end{document}